\documentclass[conference]{IEEEtran}
\IEEEoverridecommandlockouts
\usepackage{cite}
\usepackage{amsmath,amssymb,amsfonts,url}
\usepackage{algorithmic}
\usepackage{graphicx}
\usepackage{textcomp}
\usepackage{xcolor}
\usepackage{color, colortbl}
\usepackage{booktabs}
\usepackage[first=0,last=9]{lcg}
\usepackage{multirow}

\definecolor{odd}{rgb}{0.97,0.97,0.97}

\def\BibTeX{{\rm B\kern-.05em{\sc i\kern-.025em b}\kern-.08em
    T\kern-.1667em\lower.7ex\hbox{E}\kern-.125emX}}

\begin{document}

\title{Multi-Task Music Representation Learning from Multi-Label Embeddings}


\author{\IEEEauthorblockN{Alexander Schindler} 
\IEEEauthorblockA{\textit{Center for Digital Safety and Security} \\
\textit{Austrian Institute of Technology}\\
Vienna, Austria \\
alexander.schindler@ait.ac.at}
\and
\IEEEauthorblockN{Peter Knees}
\IEEEauthorblockA{\textit{Faculty of Informatics} \\
\textit{TU Wien}\\
Vienna, Austria \\
peter.knees@tuwien.ac.at}
}


\IEEEoverridecommandlockouts
\IEEEpubid{\makebox[\columnwidth]{978-1-7281-4673-7/19/\$31.00~\copyright2019 IEEE \hfill} \hspace{\columnsep}\makebox[\columnwidth]{ }}

\maketitle

\IEEEpubidadjcol

\begin{abstract}

This paper presents a novel approach to music representation learning. Triplet loss based networks have become popular for representation learning in various multimedia retrieval domains. Yet, one of the most crucial parts of this approach is the appropriate selection of triplets, which is indispensable, considering that the number of possible triplets grows cubically. We present an approach to harness multi-tag annotations for triplet selection, by using Latent Semantic Indexing to project the tags onto a high-dimensional space. From this we estimate tag-relatedness to select hard triplets. The approach is evaluated in a multi-task scenario for which we introduce four large multi-tag annotations for the Million Song Dataset for the music properties genres, styles, moods, and themes.

\end{abstract}


\section{Introduction}
\label{sec:introduction}

Search-by-example, such as finding music tracks that are similar to a query track, is an actively researched task \cite{knees2015music,knees2016music}. Research on music similarity estimation currently faces two major obstacles. First, music similarity is a highly subjective concept and is strongly influenced by the listening habits and music taste of the listener \cite{berenzweig2004large}. Second, state-of-the-art approaches to music similarity estimation are still not able to satisfactorily close the semantic gap between the computational description of music content and the perceived music similarity \cite{knees2013survey}.
The many facets of music similarity - such as specific music characteristics (e.g. rhythm, tempo, key, melody, instrumentation), perceived mood (e.g. calm, aggressive, happy), listening situation (e.g. for dinner, to concentrate, for work out), musicological factors (e.g. composer influenced by) - complicate the definition of a unified music representation which captures all semantic concepts of music.

One of the main challenges in music similarity estimation is the definition of appropriate music content descriptors to efficiently calculate similarity as a function of vector similarity or dissimilarity. Traditionally this has been approached by defining a set of features, which extract certain low level music characteristics such as timbre \cite{logan2001music} or rhythm \cite{lidy2005evaluation}, mid-level properties such as chords \cite{muller2015fundamentals}, but also high-level features. This approach faces the problem that hand-crafted feature design is neither scale-able nor sustainable \cite{humphrey2013feature}. Representation learning using Deep Neural Networks (DNN) has been actively explored in recent years \cite{sigtia2014improved,schroff2015facenet} as an alternative to feature engineering. Although some of these approaches outperform feature-based methods, a major obstacle is their dependency on large amounts of training data. Although it has been shown that shallow DNNs have an advantage on small datasets \cite{Schindler2016} they struggle to describe the latent complexity of music concepts and do not generalize on large datasets \cite{humphrey2013feature}. 
The Million Song Dataset (MSD) \cite{Bertin-Mahieux2011} is currently the largest resource for the MIR domain and consists of one million meta-data entries and links to audio-samples. A major obstacle is the absence of adequate labels or ground-truth data for these songs. Provided labels are either unstructured such as the Last.fm Dataset \cite{Bertin-Mahieux2011}, or relatively small \cite{hu2017framework}. The largest contributed ground-truth assignments \cite{schindler2012} facilitate research only on a small set of research tasks such as automatic genre classification.

In this paper we introduce four large multi-tag ground-truth assignments for tracks of the MSD for the semantic music concepts \textit{Genre}, \textit{Style}, \textit{Mood} and \textit{Theme}. This facilitates research in multiple research tasks in the MIR domain including emotion recognition, genre classification and music similarity estimation.
Through their quantity these tag-sets further facilitate improved music representation learning. Neural networks based on siamese- or triplet-network architectures have proven to be effective for this task \cite{park2018representation} using a margin optimizing loss such as contrastive \cite{hadsell2006dimensionality} or triplet loss \cite{cheng2016person}. 
Following this approach the network learns to maximize the margin between the distances of a reference track and its positive similar example and its negative dissimilar example. This learned semantic embedding space is based on the constraint that the vector distance of the positive pair should be much smaller than the distance of the negative pair.
The challenge is to select optimal positive and negative examples for a reference track. This approach was adopted from the image retrieval domain where it was successfully applied to the task of person re-identification \cite{cheng2016person}. In that scenario triplets were selected based on identity of a person. A similar approach was taken in \cite{park2018representation} where triplets were selected based on the identity of the performing artist. The problem with that approach is, that different music artists could still have similar sounding tracks. Further, distinct characteristics such as style, mood or theme cannot be estimated from artist identities. Thus, the semantic properties of the learned representation can not be influenced.
In this paper we propose an approach to harness the semantic information of the provided tag-sets. We use Latent Semantic Indexing (LSI) to project the categorical tag-information onto a high-dimensional numerical space. Tag-relatedness can then be calculated based on the vector-distance within this vector-space which is then used to select positive and negative examples for the triplet network. By combining multiple task-related tag-sets such as \textit{Genre} and \textit{Mood} their semantic context is transferred to the learned representation via the more complex tag-relatedness function.
The main contribution of our paper can be summarized as follows:

\begin{itemize}
    
    \item A novel approach to multi-task music representation learning by defining music track relatedness as a projection of categorical data onto a continuous numerical space using Latent Semantic Indexing (LSI) and using a margin optimizing online triplet loss to learn the representation
    
    \item We introduce four large multi-tag ground truth assignments for the Million Song Dataset (MSD) for the semantic music concepts \textit{Genre}, \textit{Style}, \textit{Mood} and \textit{Theme}.
    
    \item Experimental results and evaluations to demonstrate that the described approach is able to learn improved music representation by harnessing the semantic information of multiple music related tasks.
    
    
\end{itemize}



\section{Related Work}
\label{sec:relatedwork}

\paragraph{Representation Learning (RL)}

RL using DNNs gained attention through the publication of \textit{FaceNet} \cite{schroff2015facenet} which significantly improved the state-of-the-art of face re-identification. This approach is based on global item relatedness where faces belonging to the same person are similar and dissimilar otherwise. A similar approach using the global relatedness of performing artists has been applied to music data \cite{park2018representation}. Contextualized relatedness especially in the domain of music has been used in \cite{Sandouk2016}. A similar approach to ours of estimating tag-relatedness from user-tags was taken in \cite{Sandouk2016}. Latent Dirichlet Analysis (LDA) was used to project the categorical data into a numerical space. The approach was evaluated using a siamese neural network on three smaller datasets including the MSD subset using the noisy user-generated tag-sets of the Last.fm dataset. A differentiated evaluation of the learned semantic context as provided in this paper was missing.

\paragraph{Ground-truth assignments for the MSD}

To facilitate large scale MIR experiments the Million Song Dataset (MSD) \cite{Bertin-Mahieux2011} was introduced in 2011. Initially lacking ground-truth label assignments music genres were provided in \cite{schindler2012} which were collected from \textit{AllMusic} (formerly: the All Music Guide or AMG)\footnote{\url{http://allmusic.com}}. Those were verified by creating a consensus of a combination with three additional datasets \cite{schreiber2015improving}. The data collection and processing approach of this paper is similar to \cite{schindler2012}.
Ground truth assignments for mood are provided in \cite{ccano2017music}. These tags are derived from the Last.fm Dataset \cite{Bertin-Mahieux2011}. These annotations are based on user-generated tags which are highly noisy and require extensive pre-processing. The annotations contributed by our paper have been created by music expert annotators, are based on a closed vocabulary and are thus more reliable.
AllMusic Mood labels have also been used in \cite{hu2007exploring} following a similar approach as taken in our paper but on a much smaller scale and a custom dataset.
Outside the context of this paper AllMusic tags have been used in \cite{malheiro2016classification} to create a dataset of lyrics based on Valence-Arousal model.

\section{Method}
\label{sec:method}

\begin{figure}[t]
  \centering \includegraphics[width=0.4\textwidth]{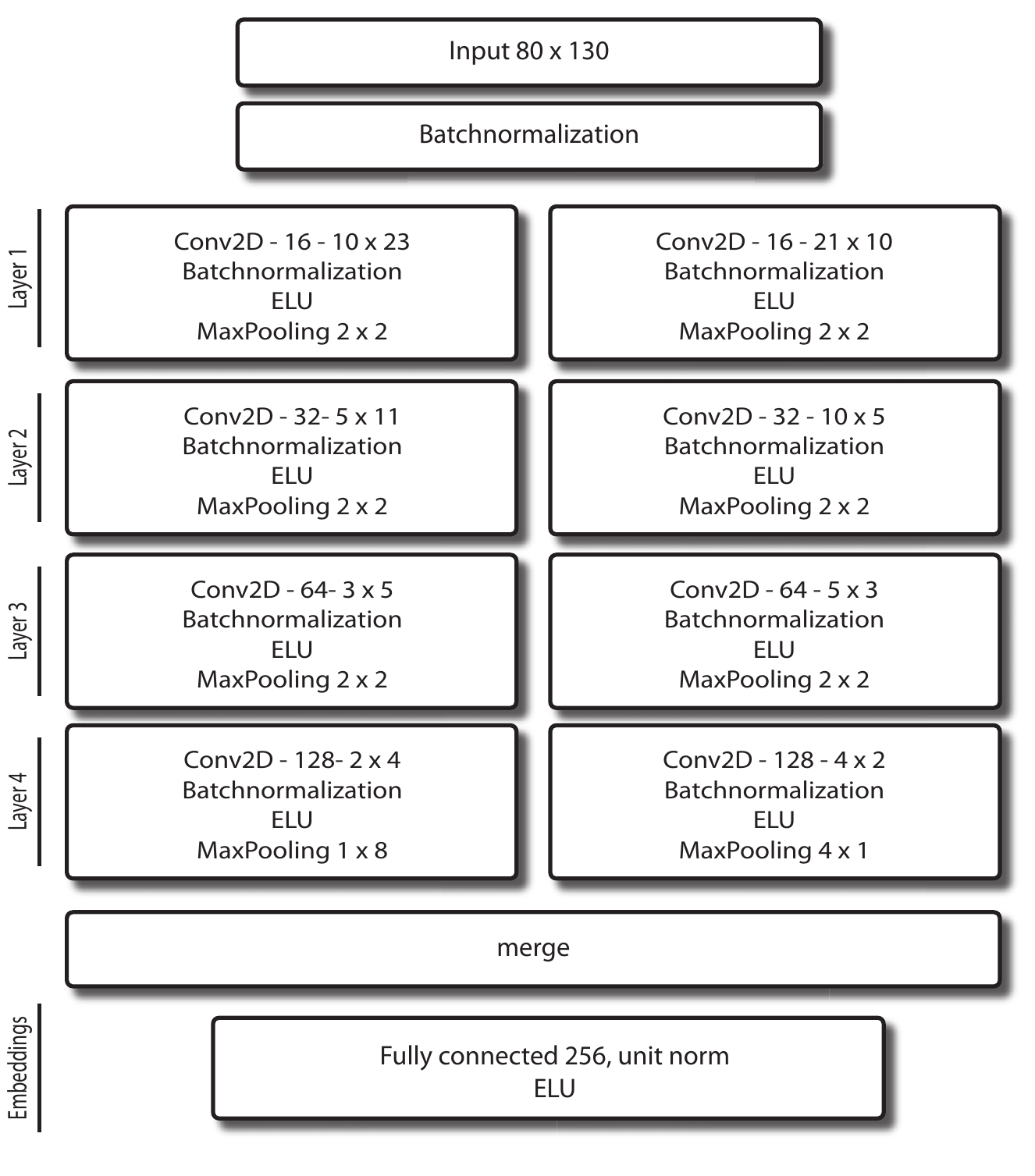}
  \caption{Deep Parallel Neural Network architecture used as base model of the triplet architecture in the evaluation. The input spectrogram is processed by two separate parallel CNN stacks to learn timbral and rhythmic features.}
  \label{fig:model_architecture}
\end{figure}


The proposed method is based on a triplet neural network architecture to learn the contextualized semantic representation using a max-margin hinge loss with online triplet selection.

\subsection{Representation Learning}

To learn the music representation we use a triplet network architecture which consists of three shared Convolutional Neural Network (CNN) stacks. The base-CNN stack is described in Section \ref{modelarchitecture} and depicted in Figure \ref{fig:model_architecture}. Using this triplet network, an input audio spectrogram $x$ is embedded $f(x_{i}^{a})$ into a  $d$-dimensional Euclidean space $\mathbb{R}^{d}$. The input consists of a triplet of music content items: a query track (anchor) $x_{i}^{a}$, a track similar (positive) $x_{i}^{p}$ and dissimilar (negative) $x_{i}^{n}$ to the anchor. The objective is to satisfy the following constraint:

\begin{equation} \label{eq1}
   \left \| f(x_{i}^{a})-  f(x_{i}^{p}) \right \|_{2}^{2} + \alpha < \left \| f(x_{i}^{a}) - f(x_{i}^{n}) \right \|_{2}^{2}  
\end{equation}

\noindent
For $\forall ( f(x_{i}^{a}), f(x_{i}^{p}), f(x_{i}^{n}) ) \in \tau $, where $\left \| f(x_{i}^{a})-  f(x_{i}^{p}) \right \|_{2}^{2}$ is the squared Euclidean distance between $x_{i}^{a}$ and $x_{i}^{p}$, which should be much smaller than the distance between $x_{i}^{a}$ and $x_{i}^{n}$. $\alpha$ is the enforced margin between positive and negative pair-distances and $\tau$ represents the set of all possible triplets in the training-set. The objective of Eq. \ref{eq1} is reformulated as the following triplet-loss function:

\begin{equation} \label{eq3}
\sum_{i=1}^{N} \max \left [ \left \| f(x_{i}^{a}) - f(x_{i}^{p}) \right \|_{2}^{2} - \left \| f(x_{i}^{a}) - f(x_{i}^{n}) \right \|_{2}^{2} + \alpha \right ]
\end{equation}

\noindent
Based on this loss function, the model learns to assign similar feature values to items with similar semantic properties and different values to dissimilar items. Efficient selection of triplets is thus a crucial step in training the network. Generating all possible triplet combinations $\tau$ is inefficient due to the cubic relation and the lacking contribution to the training-success of triplets not violating Eq. \ref{eq1}. Thus it is required to select hard triplets violating this constraint.
A common approach to this is \textit{online triplet selection} where triplets are combined within a mini-batch \cite{schroff2015facenet}. 
Our loss function takes two matrices as input: the music embeddings as provided by the model and the semantic tag-embeddings $LSI^{ts}$ (see Sect. \ref{sec:method:tagsim}).
To select positive and negative pairs, the pairwise cosine-distance $\cos(LSI_{1}^{ts}, LSI_{2}^{ts})$ matrix of the corresponding $l_{2}$ normalized semantic tag-embeddings is calculated for all mini-batch instances. The diagonal elements are set to zero to avoid identical pairs. 
For each row in this similarity matrix positive and negative pairs are selected. Thresholds for pair-selection were evaluated empirically by analyzing the distribution of the cosine-distance space of the tag-embeddings and set to $\cos(LSI_{1}^{ts}, LSI_{2}^{ts}) \geq 0.8$ (upper) and $\cos(LSI_{1}^{ts}, LSI_{2}^{ts}) < 0.2$ (lower).
For every valid pair the squared Euclidean distances of the corresponding music embeddings are calculated.
To select hard positives and negatives, $argmin$ is computed from the euclidean distances to identify the final positive and $argmax$ for the negative pair. To select appropriate triplets we use a batch size of 600 tracks. 
To avoid influences from production processes, also referred to as the ``Album Effect'' \cite{whitman2001artist}, we apply an album filter on the mini-batch to avoid pairs from the same album.
%



\subsection{Tag-Relatedness Measure}
\label{sec:method:tagsim}
\looseness -1
To present the triplet network with a target value to learn the relations between songs, we build a similarity measure based on a predefined tag set with controlled vocabulary.
We make use of labels assigned by experts, i.e. the editors of AllMusic, which is the website we extracted the labels from (cf. section~\ref{sec:dataset}).
Such expert tags typically show higher quality in annotation than user-generated tags and do not suffer from noise or vandalism~\cite{geleijnse_etal:ismir:2007,lamere:nmr:2008}.
However, especially with fine-grained genre and style taxonomies used as a basis for annotation, we find that a similarity measure based on simple tag overlap poses a problem.
For example, considering acoustically similar styles such as Punk, Punk Revival, and Pop-Punk as fully distinct categories, is not serving the purpose of learning representations optimized to resemble human perception of music. 
For track similarity calculation we therefore opt for a latent representation that takes the similarities of tags into account.

To this end, we apply \emph{latent semantic indexing} (LSI)~\cite{deerwester:lsa:1990}. 
LSI models latent topics as semantic associations of tags in a corpus of documents (in our case the tags over all music tracks).
Technically, LSI operates on the $m\times n$ weight matrix $W$ where each row corresponds to a tag $t_{j=1\ldots m}$ and each column to a music track $a_{i=1\ldots n}$.
Each cell $w_{ij} = 1$ iff tag $t\!_j$ is attributed to track $a_i$, and $w_{ij} = 0$ otherwise. 
The clustering is achieved by approximating $W$ using truncated singular value decomposition (SVD), yielding new representations of musical entities based on the uncovered topics, cf.~\cite{Knees2016}. 
As a result, each music track is represented in the LSI-derived concept space via a cluster affinity vector (in the following referred to as LSI vector).
Due to the nature of the SVD emerging topics are sorted in order of decreasing importance wrt. reconstruction of the data.
The chosen number of considered topics (i.e. the dimensionality of the LSI vectors) therefore can be used as a parameter to control generalization of the model vs preservation of the original tag information. 
\section{Evaluation}
\label{sec:evaluation}

The aim of this evaluation is to empirically asses the influence of a tag-set on the ability of a given model 
to learn a generally applicable music representation. To achieve this, the exact same model architecture as described in the following subsection is used for all experiments. The only parameter that changes in the experiments are the differently generated LSI vectors for each track. We further controlled for all random processes such as kernel initialization or shuffling of training instances after each epoch. The same training, validation and test splits are used in all experiments. By controlling all these parameters to our best knowledge we hypothesize that the learned representations are only influenced by LSI representation of the semantic space of the tag-sets and tag-set combinations.

\begin{figure*}[t]
  \centering
  \includegraphics[width=\linewidth]{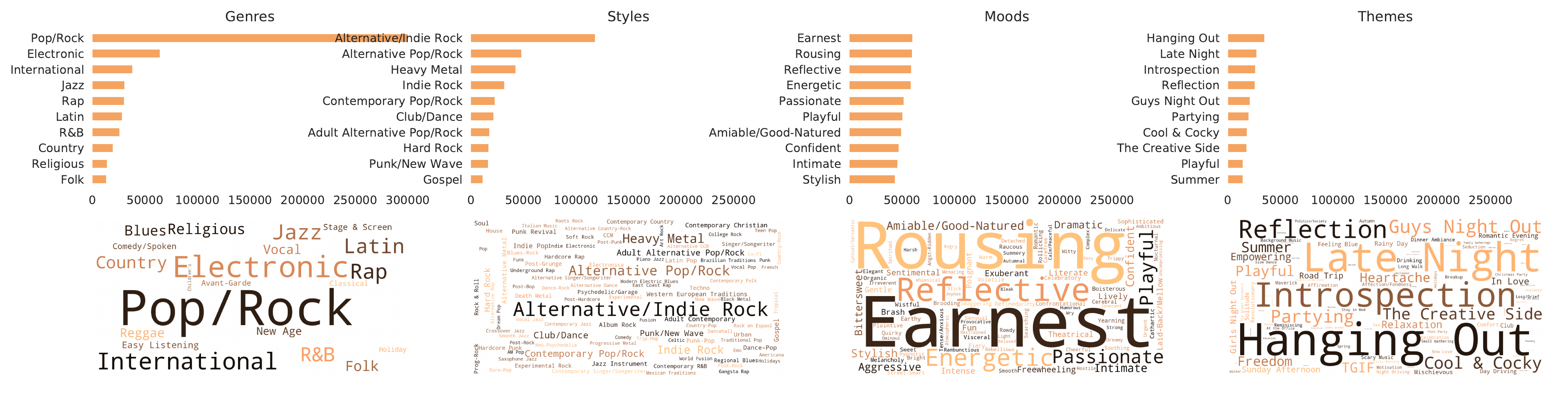}
  \caption{Tag-Set overview. Top-10 most frequent tags for each set.}
  \label{fig_tagsets}
\end{figure*}

\subsection{Model Architecture - Deep Parallel Neural Networks}
\label{modelarchitecture}

For the evaluation we are using a slight adaption of the Deep Parallel Neural Network (DPNN) architecture as described in \cite{Schindler2016}. It is a parallel arrangement of rectangular shaped filters and Max-Pooling windows to capture frequency and temporal relationships at once (see Figure \ref{fig:model_architecture}). The Parallel Neural Network (PNN) architecture was first described in \cite{pons_cbmi2016}. It generally consists of two - not shared - Convolutional Neural Network (CNN) stacks which process the same input Spectrogram in parallel. The aim is to learn intermediate embeddings for rhythm and timbre separately which are then concatenated to a combined music representation. To achieve this, the left stack aims at capturing frequency relations using horizontally aligned rectangular filter kernel shapes \cite{lidy2016cqt}. By applying Max Pooling to reduce information on the temporal axis, this stack is forced to facilitate spectral information. The right stack performs adequately but uses vertically aligned rectangular filter shapes and applies Max Pooling to reduce information on the spectral axis. Its aim is to capture rhythmic patterns over time.
Due to the large number of data instances in this evaluation, we are using the Deep Parallel Neural Network architecture \cite{Schindler2016} which showed better performance on larger data-sets. This architecture is further adapted by applying Batch normalization to each convolution layer and by using Exponential Linar Units (ELU) as activation function for all layers. Due to the larger input Spectrogram, the Max Pooling layers were also slightly different parametrized (see Figure \ref{fig:model_architecture}).
The final representation layer has a dimensionality of 256.

\begin{figure*}[t]
  \centering
  \includegraphics[scale=0.44]{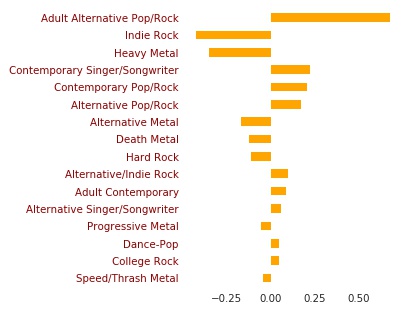} 
  \includegraphics[scale=0.44]{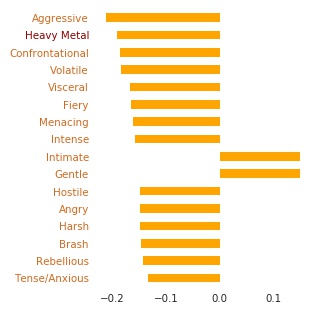} 
  \includegraphics[scale=0.44]{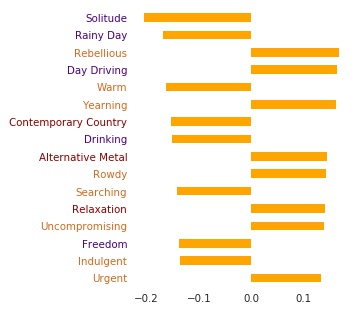} 
  \includegraphics[scale=0.44]{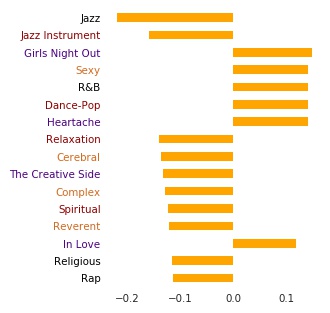}
  \caption{Exemplary LSI topics from mixtures of tag sets (topic numbers in parentheses). From left to right: styles (\#2), styles+moods (\#1), styles+moods+themes (\#61), and styles+moods+themes+genres (\#7). Styles in red, moods in orange, themes in purple, genres in black.}
  \label{fig:lsi_topics}
\end{figure*}

\subsection{Data Pre-processing}
\label{dataprep}

\subsubsection{Tag-Set Pre-processing}

To avoid influences through sparsity within the annotation spaces, we only use the intersection of the tag-sets where for each track and tag-set at least one tag and an audio sample is available. This step reduced the size of the data-set to about 58\% on average and eliminated 106 unique style and one mood tags (see Table \ref{tab:tags_overview}).

\subsubsection{Audio Data Processing}

The evaluation uses audio samples assembled in \cite{schindler2012} which are re-sampled to 22.050 Hz. A 6 seconds segment is read using a 3 seconds offset to avoid frequent fade-in effects. Short-time Fourier Transform (STFT) with a Hanning-windowing function and a 2048 samples window size with 50\% overlap is applied. The Spectrograms are transformed to the log-transformed Mel-space using 80 Mel-filters and cut-off frequencies of 16Hz (min) and 11.000Hz (max) resulting in an input matrix shape of 80x130x1. Instead of normalizing the feature-space, we add a batch-normalization layer on top of the neural network (see Fig. \ref{fig:model_architecture}).

\section{Tag-Set Collections}
\label{sec:dataset}

The Million Song Dataset (MSD) is currently the largest music dataset \cite{Bertin-Mahieux2011} and provides meta-data for one million contemporary (until 2011) songs, including attributes such as titles, artist and album names, but also references to third party meta-data repositories such as MusicBrainz or 7Digital over which audio samples can be obtained. AllMusic \cite{datta2002} is a Web portal of a large music information database including album reviews, artist biographies as well as expert tagging for albums according into genres, styles, moods and themes. Data-collection is aligned to \cite{schindler2012}. Meta-data was automatically collected from AllMusic using a web-scraping script based on direct string matching to query for artist-release combinations. From the resulting Album Web page genre, style, mood and theme tags were collected (see Table \ref{tab:tags_overview}). 
The \textit{Genres Tag-Set} is the largest collection with a skewed distribution towards the label 'Pop/Rock'. The set further contains cross-genre tags such as \textit{Holiday} and \textit{Children} (see Fig. \ref{fig_tagsets}).
The \textit{Styles Tag-Set} complement genres and provide a more detailed description of music characteristic such as rhythm, harmony, instrumentation, etc. 
Tags of the \textit{Moods Tag-Set} depict emotions expressed by the music.
The \textit{Themes Tag-Set} describes certain cross-genre scenarios or occasions such as holiday, party or Christmas.
\textit{The introduced multi-label tag-sets are provided for download from our MSD-benchmark webpage: \url{http://www.ifs.tuwien.ac.at/mir/msd/download.html}}.

\begin{table}[t]
\centering
\caption{Tag-Set statistical overview.}
\begin{tabular}{l|rrrr}
\toprule
{} &  \textbf{Genres} &  \textbf{Styles} &   \textbf{Moods} &  \textbf{Themes} \\
\midrule
\rowcolor{odd}
\cellcolor{white!} Unique Tags      &      21 &     939 &     286 &     166 \\
\cellcolor{white!} Tag Combinations &     688 &   13589 &   22577 &    7322 \\
\rowcolor{odd}
\cellcolor{white!} Labelled Albums  &   75339 &   52304 &   32148 &   19375 \\
\cellcolor{white!} Labelled Tracks  &  504502 &  364326 &  229510 &  145555 \\
\midrule
\rowcolor{odd}
\cellcolor{white!} Unique Tags      &      21 &     833 &     285 &     166 \\
\cellcolor{white!} Tag Combinations &     449 &    7446 &   14300 &    7298 \\
\rowcolor{odd}
\cellcolor{white!} Labelled Albums  &   19107 &   19107 &   19107 &   19107 \\
\cellcolor{white!} Labelled Tracks  &  143587 &  143587 &  143587 &  143587 \\
\bottomrule
\end{tabular}

\label{tab:tags_overview}
\end{table}

\subsection{LSI-Transformed Topic Sets}

To avoid the issues coming from an overlap-oriented similarity measure described in section~\ref{sec:method:tagsim}, we apply LSI to model latent topics.
We further want to investigate the potential of modeling joint representations of tags from different learning tasks. 
To this end, beside modeling individual tag sets (see above) using LSI, we also exhaustively explore combinations of two, three, and four tag sets from the different tasks by joining tag sets prior to applying LSI.
Figure~\ref{fig:lsi_topics} shows exemplary resulting LSI topics from different mixtures of tag sets. 
The number in parenthesis refers to ordinal number of the concept cluster as obtained from the SVD process.
As the topics are connected to the singular values which are ordered in descending relevance, the ordinal number of the cluster is indicative of the relevance in describing the data.
For instance, for LSI topic \#2 derived from style tags only, we can see how concept modeling a distinction between various types of alternative and contemporary rock and songwriting on one hand (positive loading), and various forms of heavy metal on the other (negative loading).
For the \#1 topic derived from a combination of styles and mood tags we can see a separation of the moods ``intimate'' and ``gentle'' from many mood variants capturing ``aggressive'' and ``angry'' sentiment.
While this concept is heavily defined by moods, the style heavy metal contributes to the aggressive side.
The combined styles, moods, and themes topic \#61 (as an example of a higher order concept capturing more fine-grained aspects) shows more introverted facets on one side (with themes like ``solitude'', ``rainy day'', and ``drinking''; ``contemporary country'' as a style; and moods like ``warm'', ``searching'', and ``indulgent'') as opposed to urgent activity on the other (with themes like ``day driving''; styles like ``alternative metal''; and moods like ``rebellious'', ``yearning'', ``rowdy'', and ``uncompromising'').
Finally, from a combination of styles, moods, themes, and genres, we see how conceptual clusters span across tasks and tag sets. 
As such, topic \#7 connects themes like ``girls night out'', ``heartache'', and ``in love'' with ``dance pop'' (style), ``sexy'' (mood), and ``r\&b'' (genre).
This is opposed by two trends, namely musical sophistication (``jazz'', ``cerebral'', ``creative'', ``complex'') and spiritual music (``spiritual'', ``religious'').


\section{Results}

Experimental results are provided in Table \ref{tab:results}. To guarantee that the semantic concepts of the tag-sets is the only influencing factor to the results, we took care to control every random process in the experiments including kernel initialization of the networks layers, as well the shuffling of training instances after each epoch. We used constant artist-stratified train, validation and test splits (122.766, 6.461, 14.358). 
Experiments were performed by increasing the number of LSI topics by steps of 10 until 400 for each tag-set as well for each tag-set combination. The learned representations were evaluated in different tasks according their precision (at cut-off 100) to retrieve tracks of the same \textit{genre}, \textit{style}, \textit{mood}, \textit{theme}, \textit{artist} or \textit{album}.
The results show that the coarse defined \textit{genre} tag-set only contributes to the retrieval of itself and does not gain from combinations with other tags.
It is interesting to observe, that representation learned with the \textit{mood} tag-set show much higher precision values than \textit{style} tags, which are more frequent.
This can also be observed for all tag-set combinations. All combinations containing \textit{mood} tags outperform the others.
Further, it can be observed that the latent semantic information is better captured using more LSI concepts - although we observed that there is a glass ceiling.

\section{Conclusions and Future Work}

We introduced a novel approach to music representation learning. We showed how to estimate tag-relatedness from multi-tag annotations by projecting the categorical data into a vector space using Latent Semantic Indexing. This tag-relatedness is used for online triplet selection to train a triplet deep neural network. This approach solved two issues. First, how to estimate subjective music similarity for triplet selection and second, how to estimate tag-relatedness from multi-label annotations. In the experiments we showed that this method facilitates to learn music representations which are optimized towards the semantic context of a given task. 
Future work could extend the track relatedness to other modalities such as song lyrics, album reviews or album-cover arts.

\begin{table*}[t]
\caption{Experimental results (in Precision at cut-off 100) for single tasks, two, three and four tasks tag-set combinations.}
\begin{tabular}{c|lcrrrrrr}
\toprule
 & \textbf{Tag-Set} &  \textbf{LSI Topics} &  \textbf{Prec. Genres} &  \textbf{Prec. Styles} &  \textbf{Prec. Moods} &  \textbf{Prec. Themes} &  \textbf{Prec. Artists} &  \textbf{Prec. Album} \\
\midrule

\rowcolor{odd}
 \cellcolor{white!} & genres &  10 & 0.3951 & 0.0091 & 0.0060 & 0.0076 & 0.0085 & 0.0047 \\
 \cellcolor{white!} & genres &  3 & \textbf{0.3971} & 0.0082 & 0.0055 & 0.0070 & 0.0078 & 0.0043 \\
 \rowcolor{odd}
 \cellcolor{white!} & moods &  160 & 0.3597 & \textbf{0.0148} & \textbf{0.0101} & \textbf{0.0122} & \textbf{0.0143} & \textbf{0.0078} \\
  & moods &  20 & 0.3265 & 0.0100 & 0.0066 & 0.0084 & 0.0095 & 0.0050 \\
 \rowcolor{odd}
 \cellcolor{white!} & styles &  200 & 0.3724 & 0.0129 & 0.0079 & 0.0100 & 0.0116 & 0.0061 \\
 \cellcolor{white!} & styles &  20 & 0.3586 & 0.0100 & 0.0063 & 0.0081 & 0.0092 & 0.0047 \\
 \rowcolor{odd}
 \cellcolor{white!} & themes &  60 & 0.3160 & 0.0082 & 0.0052 & 0.0067 & 0.0074 & 0.0039 \\
\multirow{-8}{*}{\rotatebox[origin=c]{90}{\textbf{Single-Task}}}  & themes &  20 & 0.3005 & 0.0077 & 0.0050 & 0.0064 & 0.0072 & 0.0038 \\
 \midrule
 \rowcolor{odd}
 \cellcolor{white!} & genres-moods &  180 & 0.3550 & 0.0154 & 0.0106 & 0.0127 & 0.0149 & 0.0082 \\
 \cellcolor{white!} & genres-moods &  20 & 0.3415 & 0.0105 & 0.0069 & 0.0089 & 0.0099 & 0.0052 \\
 \rowcolor{odd}
 \cellcolor{white!} & genres-styles &  200 & \textbf{0.3802} & 0.0132 & 0.0082 & 0.0106 & 0.0120 & 0.0062 \\
 \cellcolor{white!} & genres-styles &  20 & 0.3633 & 0.0104 & 0.0066 & 0.0085 & 0.0094 & 0.0050 \\
 \rowcolor{odd}
 \cellcolor{white!} & genres-themes &  80 & 0.3539 & 0.0103 & 0.0065 & 0.0086 & 0.0096 & 0.0050 \\
 \cellcolor{white!} & genres-themes &  20 & 0.3279 & 0.0092 & 0.0060 & 0.0077 & 0.0085 & 0.0046 \\
 \rowcolor{odd}
 \cellcolor{white!} & moods-themes &  220 & 0.3624 & \textbf{0.0161} & \textbf{0.0114} & \textbf{0.0136} & 0.0157 & 0.0088 \\
 \cellcolor{white!} & moods-themes &  20 & 0.3304 & 0.0105 & 0.0068 & 0.0087 & 0.0098 & 0.0051 \\
 \rowcolor{odd}
 \cellcolor{white!} & styles-moods &  320 & 0.3660 & \textbf{0.0161} & \textbf{0.0114} & \textbf{0.0136} & \textbf{0.0160} & \textbf{0.0089} \\
 \cellcolor{white!} & styles-moods &  20 & 0.3423 & 0.0114 & 0.0074 & 0.0095 & 0.0107 & 0.0056 \\
 \rowcolor{odd}
 \cellcolor{white!} & styles-themes &  240 & 0.3551 & 0.0145 & 0.0095 & 0.0119 & 0.0136 & 0.0074 \\
 \multirow{-12}{*}{\rotatebox[origin=c]{90}{\textbf{2-Tasks}}} & styles-themes &  20 & 0.3418 & 0.0108 & 0.0068 & 0.0087 & 0.0099 & 0.0051 \\
\midrule
 \rowcolor{odd}
 \cellcolor{white!} & genres-moods-themes &  200 & 0.3679 & 0.0159 & 0.0113 & 0.0134 & 0.0156 & 0.0087 \\
 \cellcolor{white!} & genres-moods-themes &  20 & 0.3377 & 0.0109 & 0.0069 & 0.0089 & 0.0101 & 0.0052 \\
 \rowcolor{odd}
 \cellcolor{white!} & genres-styles-moods &  400 & 0.3708 & \textbf{0.0173} &\textbf{ 0.0122} & \textbf{0.0145} & \textbf{0.0170} & \textbf{0.0094} \\
 \cellcolor{white!} & genres-styles-moods &  20 & 0.3606 & 0.0121 & 0.0077 & 0.0100 & 0.0112 & 0.0059 \\
 \rowcolor{odd}
 \cellcolor{white!} & genres-styles-themes &  200 & \textbf{0.3724} & 0.0145 & 0.0097 & 0.0120 & 0.0138 & 0.0074 \\
 \cellcolor{white!} & genres-styles-themes &  20 & 0.3616 & 0.0114 & 0.0070 & 0.0091 & 0.0102 & 0.0052 \\
 \rowcolor{odd}
 \cellcolor{white!} & styles-moods-themes &  360 & 0.3602 & 0.0170 & 0.0119 & 0.0142 & 0.0166 & 0.0093 \\
 \multirow{-8}{*}{\rotatebox[origin=c]{90}{\textbf{3-Tasks}}} & styles-moods-themes &  20 & 0.3399 & 0.0113 & 0.0073 & 0.0092 & 0.0106 & 0.0055 \\
\midrule
 \rowcolor{odd}
 \cellcolor{white!} & genres-styles-moods-themes &  340 & \textbf{0.3652} & \textbf{0.0173} & \textbf{0.0125} & \textbf{0.0147} & \textbf{0.0173} & \textbf{0.0097} \\
 \multirow{-2}{*}{\rotatebox[origin=c]{90}{\textbf{4-T.}}} & genres-styles-moods-themes &  20 & 0.3551 & 0.0117 & 0.0076 & 0.0097 & 0.0112 & 0.0058 \\
\bottomrule
\end{tabular}
\label{tab:results}
\end{table*}

\medskip
\small
\textbf{Acknowledgements} \textit{This article has been made possible partly by received funding from the European Unions Horizon 2020 research and innovation program in the context
of the VICTORIA project under grant agreement no. SEC-740754}

\bibliography{references}

\begin{thebibliography}{10}
\providecommand{\url}[1]{#1}
\csname url@samestyle\endcsname
\providecommand{\newblock}{\relax}
\providecommand{\bibinfo}[2]{#2}
\providecommand{\BIBentrySTDinterwordspacing}{\spaceskip=0pt\relax}
\providecommand{\BIBentryALTinterwordstretchfactor}{4}
\providecommand{\BIBentryALTinterwordspacing}{\spaceskip=\fontdimen2\font plus
\BIBentryALTinterwordstretchfactor\fontdimen3\font minus
  \fontdimen4\font\relax}
\providecommand{\BIBforeignlanguage}[2]{{%
\expandafter\ifx\csname l@#1\endcsname\relax
\typeout{** WARNING: IEEEtran.bst: No hyphenation pattern has been}%
\typeout{** loaded for the language `#1'. Using the pattern for}%
\typeout{** the default language instead.}%
\else
\language=\csname l@#1\endcsname
\fi
#2}}
\providecommand{\BIBdecl}{\relax}
\BIBdecl

\bibitem{knees2015music}
P.~Knees and M.~Schedl, ``Music retrieval and recommendation: A tutorial
  overview,'' in \emph{Proc. Int. ACM SIGIR Conf. on Research and Development
  in Information Retrieval}, 2015.

\bibitem{knees2016music}
------, \emph{Music similarity and retrieval: An introduction to audio-and
  web-based strategies}.\hskip 1em plus 0.5em minus 0.4em\relax Springer, 2016,
  vol.~36.

\bibitem{berenzweig2004large}
A.~Berenzweig, B.~Logan, D.~P. Ellis, and B.~Whitman, ``A large-scale
  evaluation of acoustic and subjective music-similarity measures,''
  \emph{Computer Music Journal}, vol.~28, no.~2, pp. 63--76, 2004.

\bibitem{knees2013survey}
P.~Knees and M.~Schedl, ``A survey of music similarity and recommendation from
  music context data,'' \emph{ACM Transactions on Multimedia Computing,
  Communications, and Applications (TOMM)}, 2013.

\bibitem{logan2001music}
B.~Logan and A.~Salomon, ``A music similarity function based on signal
  analysis.'' in \emph{ICME}, 2001, pp. 22--25.

\bibitem{lidy2005evaluation}
T.~Lidy and A.~Rauber, ``Evaluation of feature extractors and psycho-acoustic
  transformations for music genre classification,'' in \emph{ISMIR}, 2005.

\bibitem{muller2015fundamentals}
M.~M{\"u}ller, \emph{Fundamentals of music processing: Audio, analysis,
  algorithms, applications}.\hskip 1em plus 0.5em minus 0.4em\relax Springer,
  2015.

\bibitem{humphrey2013feature}
E.~J. Humphrey, J.~P. Bello, and Y.~LeCun, ``Feature learning and deep
  architectures: New directions for music informatics,'' \emph{Journal of
  Intelligent Information Systems}, vol.~41, no.~3, pp. 461--481, 2013.

\bibitem{sigtia2014improved}
S.~Sigtia and S.~Dixon, ``Improved music feature learning with deep neural
  networks,'' in \emph{2014 IEEE international conference on acoustics, speech
  and signal processing (ICASSP)}.\hskip 1em plus 0.5em minus 0.4em\relax IEEE,
  2014, pp. 6959--6963.

\bibitem{schroff2015facenet}
F.~Schroff, D.~Kalenichenko, and J.~Philbin, ``Facenet: A unified embedding for
  face recognition and clustering,'' in \emph{Proceedings of the IEEE
  conference on computer vision and pattern recognition}, 2015.

\bibitem{Schindler2016}
A.~Schindler, T.~Lidy, and A.~Rauber, ``{Comparing shallow versus deep neural
  network architectures for automatic music genre classification},'' in
  \emph{9th Forum Media Technology (FMT2016)}, vol. 1734, 2016, pp. 17--21.

\bibitem{Bertin-Mahieux2011}
T.~Bertin-Mahieux, D.~P. Ellis, B.~Whitman, and P.~Lamere, ``The million song
  dataset,'' in \emph{Proceedings of the International Conference on Music
  Information Retrieval (ISMIR2011)}, vol.~2, no.~9, 2011, p.~10.

\bibitem{hu2017framework}
X.~Hu, K.~Choi, and J.~S. Downie, ``A framework for evaluating multimodal music
  mood classification,'' \emph{Journal of the Association for Information
  Science and Technology}, vol.~68, no.~2, pp. 273--285, 2017.

\bibitem{schindler2012}
A.~Schindler, R.~Mayer, and A.~Rauber, ``Facilitating comprehensive
  benchmarking experiments on the million song dataset,'' in \emph{Proc. of the
  13th Int. Conf. on Music Information Retrieval (ISMIR 2012)}, 2012.

\bibitem{park2018representation}
J.~Park, J.~Lee, J.~Park, J.-W. Ha, and J.~Nam, ``Representation learning of
  music using artist labels,'' in \emph{19th International Society for Music
  Information Retrieval Conference (ISMIR 2018)}, 2018.

\bibitem{hadsell2006dimensionality}
R.~Hadsell, S.~Chopra, and Y.~LeCun, ``Dimensionality reduction by learning an
  invariant mapping,'' in \emph{2006 IEEE Computer Society Conference on
  Computer Vision and Pattern Recognition (CVPR'06)}, vol.~2.\hskip 1em plus
  0.5em minus 0.4em\relax IEEE, 2006, pp. 1735--1742.

\bibitem{cheng2016person}
D.~Cheng, Y.~Gong, S.~Zhou, J.~Wang, and N.~Zheng, ``Person re-identification
  by multi-channel parts-based cnn with improved triplet loss function,'' in
  \emph{Proceedings of the IEEE Conference on Computer Vision and Pattern
  Recognition}, 2016, pp. 1335--1344.

\bibitem{Sandouk2016}
U.~Sandouk and K.~Chen, ``{Learning Contextualized Music Semantics from Tags
  Via a Siamese Neural Network},'' \emph{ACM Transactions on Intelligent
  Systems and Technology}, vol.~8, no.~2, pp. 1--21, 2016.

\bibitem{schreiber2015improving}
H.~Schreiber, ``Improving genre annotations for the million song dataset.'' in
  \emph{ISMIR}, 2015, pp. 241--247.

\bibitem{ccano2017music}
E.~{\c{C}}ano and M.~Morisio, ``Music mood dataset creation based on last. fm
  tags,'' in \emph{2017 International Conference on Artificial Intelligence and
  Applications, Vienna, Austria}, 2017.

\bibitem{hu2007exploring}
X.~Hu and J.~S. Downie, ``Exploring mood metadata: Relationships with genre,
  artist and usage metadata.'' in \emph{International Conference on Artificial
  Intelligence and Applications (ISMIR2007)}, 2007, pp. 67--72.

\bibitem{malheiro2016classification}
R.~Malheiro, R.~Panda, P.~Gomes, and R.~P. Paiva, ``Classification and
  regression of music lyrics: Emotionally-significant features,'' in \emph{8th
  Int. Conf. on Knowledge Discovery and Information Retrieval}, 2016.

\bibitem{whitman2001artist}
B.~Whitman, G.~Flake, and S.~Lawrence, ``Artist detection in music with
  minnowmatch,'' in \emph{Neural Networks for Signal Processing XI: Proceedings
  of the 2001 IEEE Signal Processing Society Workshop (IEEE Cat. No.
  01TH8584)}.\hskip 1em plus 0.5em minus 0.4em\relax IEEE, 2001, pp. 559--568.

\bibitem{geleijnse_etal:ismir:2007}
G.~Geleijnse, M.~Schedl, and P.~Knees, ``{The Quest for Ground Truth in Musical
  Artist Tagging in the Social Web Era},'' in \emph{Proceedings of the
  8\textsuperscript{th} International Conference on Music Information Retrieval
  {(ISMIR)}}, Vienna, Austria, 2007.

\bibitem{lamere:nmr:2008}
P.~Lamere, ``{Social Tagging and Music Information Retrieval},'' \emph{{Journal
  of New Music Research}}, vol.~37, no.~2, pp. 101--114, 2008.

\bibitem{deerwester:lsa:1990}
S.~Deerwester, S.~T. Dumais, G.~W. Furnas, T.~K. Landauer, and R.~Harshman,
  ``Indexing by latent semantic analysis,'' \emph{Journal of the American
  Society for Information Science}, vol.~41, pp. 391--407, 1990.

\bibitem{Knees2016}
P.~Knees and M.~Schedl, ``Contextual music similarity, indexing, and
  retrieval,'' in \emph{Music Similarity and Retrieval: An Introduction to
  Audio- and Web-based Strategies}.\hskip 1em plus 0.5em minus 0.4em\relax
  Berlin, Heidelberg: Springer Berlin Heidelberg, 2016, pp. 133--158.

\bibitem{pons_cbmi2016}
J.~Pons, T.~Lidy, and X.~Serra, ``Experimenting with musically motivated
  convolutional neural networks,'' in \emph{Proceedings of the 14th
  International Workshop on Content-based Multimedia Indexing (CBMI 2016)},
  Bucharest, Romania, June 2016.

\bibitem{lidy2016cqt}
T.~Lidy and A.~Schindler, ``Cqt-based convolutional neural networks for audio
  scene classification,'' in \emph{Proceedings of the Detection and
  Classification of Acoustic Scenes and Events 2016 Workshop (DCASE2016)},
  vol.~90.\hskip 1em plus 0.5em minus 0.4em\relax DCASE2016 Challenge, 2016,
  pp. 1032--1048.

\bibitem{datta2002}
D.~Datta, ``Managing metadata,'' in \emph{Proc. of the International Conference
  on Music Information Retrieval (ISMIR2002)}, Paris, France, Oct. 2002.

\end{thebibliography}
\bibliographystyle{IEEEtran}

\end{document}